\documentclass[aps,prb,twocolumn,floatfix,amsmath,amssymb,showpacs,
               superscriptaddress]{revtex4-1}
\pdfoutput=1
\usepackage{graphicx,placeins}
\usepackage{color}
\usepackage{dcolumn}
\usepackage{bm}
\usepackage{epsfig,psfrag,amsmath,amssymb,float}
\input{epsf}
\usepackage{hyperref}
\usepackage[percent]{overpic}
\hypersetup{
    colorlinks=true,
    linkcolor=blue,
    citecolor=blue,
    filecolor=magenta,      
    urlcolor=cyan,
}
\usepackage{array}
\newcolumntype{C}[1]{>{\centering\arraybackslash$}p{#1}<{$}}

\setcitestyle{square}
\newcommand{\pdag}{{\phantom{\dagger}}}
\bibpunct{[}{]}{,}{n}{}{}
\begin{document}
\title{Magnetization dynamics fingerprints of an excitonic condensate $t_{2g}^{4}$ magnet}

\author{Nitin Kaushal}
\affiliation{Materials Science and Technology Division, Oak Ridge National 
Laboratory, Oak Ridge, Tennessee 37831, USA}

\author{Jacek Herbrych}
\affiliation{Department of Theoretical Physics, Faculty of Fundamental Problems of Technology%
, Wroc\l{}aw University of Science and Technology, 50-370 Wroc\l{}aw, Poland}

\author{Gonzalo Alvarez}
\affiliation{Materials Science and Technology Division, Oak Ridge National 
Laboratory, Oak Ridge, Tennessee 37831, USA}
\affiliation{Computational Sciences and Engineering Division %
and Center for Nanophase Materials Sciences, Oak Ridge National Laboratory, %
Oak Ridge, Tennessee 37831, USA}

\author{Elbio Dagotto}
\affiliation{Materials Science and Technology Division, Oak Ridge National 
Laboratory, Oak Ridge, Tennessee 37831, USA}
\affiliation{Department of Physics and Astronomy, The University of 
Tennessee, Knoxville, Tennessee 37996, USA}

\date{\today}

\begin{abstract}
The competition between spin-orbit coupling $\lambda$ and electron-electron interaction $U$ leads to a plethora of novel states of matter, extensively studied in the context of $t_{2g}^4$ and $t_{2g}^5$ materials, such as ruthenates and iridates. Excitonic magnets -- the antiferromagnetic state of bounded electron-hole pairs -- is a prominent example of phenomena driven by those competing energy scales. Interestingly, recent theoretical studies predicted that excitonic magnets can be found in the ground-state of spin-orbit-coupled $t_{2g}^4$ Hubbard models. Here, we present a detailed computational study of the magnetic excitations in that excitonic magnet, employing one-dimensional chains (via density matrix renormalization group) and small two-dimensional clusters (via Lanczos).  Specifically, first we show that the low-energy spectrum is dominated by a dispersive (acoustic) magnonic mode, with extra features arising from the $\lambda=0$ state in the phase diagram. 
Second, and more importantly, we found a novel magnetic excitation forming a high-energy optical mode with the highest intensity at wavevector $q\to 0$. In
the excitonic condensation regime at large $U$, we also have found a novel high-energy $\pi$-mode composed solely
of orbital excitations.
These unique fingerprints of the $t_{2g}^4$ excitonic magnet are important in the analysis of neutron and RIXS experiments.

\end{abstract}
\maketitle

\section{Introduction}

The search for conclusive experimental evidence for an excitonic condensate is a challenging task.
Although the mathematical close similarity betweeen BCS superconductors~\cite{Bardeen01} and 
excitonic condensates~\cite{rice1960,rice1960-1,rice1960-2,Mott01,Knox01} became apparent since early 
theoretical descriptions, superconductors have simple experimental ``smoking guns'': zero resistivity and
Meissner effect. However, thus far only a handful of experimental studies, such as momentum-resolved electron energy-loss spectroscopy of 1T-TiSe$_{2}$~\cite{Kogar01}, transport and optical conductivity measurements of quasi-one-dimensional Ta$_{2}$NiSe$_{5}$~\cite{Larkin01,YFLu01,Sugimoto01}, vanishing Hall resistance in bilayer electron systems~\cite{Eisenstein01}, and tunneling experiments between graphene bilayers~\cite{Burg01} have been interpreted as providing evidence of excitonic condensation. 

Interestingly, the condensation of spinfull excitons can also lead to the development of magnetism in the condensate~\cite{Kunes01,Kaneko02,Kaushal02}. Recently, $4d/5d$ transition metal oxides~\cite{Krempa01,Cao02,Rau01}, because of their robust spin-orbit coupling (SOC), have become common playgrounds to search for the predicted exciton magnets in real materials~\cite{Dasgupta01}.
 The delicate interplay between various energy scales in these compounds, i.e. robust SOC, Coulomb electron-electron interations, and the kinetic energy of electrons, may drive condensation of spin-orbit excitons (electron-hole bound states in $j_{\textrm{eff}}=1/2$ and $j_{\textrm{eff}}=3/2$ bands). These excitons manifest as an atomic bound state involving $J=1$ (triplons) in the large Hubbard $U$ limit. 
Their condensation leads to antiferromagnetic ordering, hence named excitonic magnet~\cite{Khaliullin}. In this context, exotic magnetic states have been recently analyzed in $t_{2g}^{4}$ spin-orbit coupled multi-orbital systems via various theoretical and computational studies~\cite{Khaliullin,meetei,Svoboda01,Kaushal01,Kaushal03,Sato02,exc2}. The presence of antiferromagnetic correlations and the mechanism behind their development were recently under debate also in the double perovskite iridates Sr$_2$YIrO$_6$, Ba$_2$YIrO$_6$ ($5d^4$)~\cite{Cao01,Bhowal01,Nag01,Dey01,Pajskr01,Hammermath01,Chen01,Fuchs01,Gong01,Terzic01,Corredor01,Phelan02,Ranjbar01} However, resonant inelastic X-ray scattering (RIXS) experiments~\cite{Kusch01} and theoretical studies~\cite{BHKim01} showed that the SOC is large enough to gap the triplon excitations, hence the excitonic condensation is unlikely to occur in these materials. Furthermore, other Ir-based compounds, such as Ba$_3$Zn(Y)Ir$_2$O$_9$ ~\cite{Nag02,Nag03,Dey02}, Ba(Sr)ScIrO$_6$~\cite{Chakraborty01}, and SrLaNiIrO$_6$~\cite{Wolff01}, have displayed magnetism, or at least weak moments, instead of the expected $J=0$ non-magnetic state.

An intensively studied quasi two-dimensional $4d^{4}$ compound Ca$_2$RuO$_4$ also shows antiferromagnetism below $T_{N}\approx 110K$~\cite{Braden01}. Recently, inelastic neutron scattering (INS) results on Ca$_2$RuO$_4$~\cite{Jain01} displayed a gapped low-energy transverse spin-wave branch with XY-like dispersion and another gapped higher-energy amplitude branch, as moments are confined to the $a-b$ plane. Both branches were explained by incorporating spin-orbit coupling and tetragonal crytal-field splitting in spin-wave calculations using the phenomenological singlet-triplet model~\cite{Akbari01} which captures the triplon condensation. However, it is worth noting that the microscopic understanding of the magnetism in Ca$_2$RuO$_4$ is still highly debatable. Recent local-density approximation + dynamical mean field theory calculations have indicated large $xy$-orbital polarization ($n_{xy} \approx 2$ and $n_{xz}\approx n_{yz} \approx 1$) due to tetragonal splitting and the above result is nearly unaffected by SOC~\cite{Zhang01}. These findings suggest that Ca$_2$RuO$_4$ could be a conventional $S=1$ Mott insulator. The low-energy in-plane transverse mode of INS has also been described using the $S=1$ Heisenberg model with single-ion anisotropy (coming from SOC) via spin-wave theory~\cite{Kunkemoller01}. These results suggest that SOC, though important, is not a decisive component to explain the ground state of Ca$_2$RuO$_4$. Moreover, reducing the flattening of the RuO$_4$ octahedra by applying pressure leads to a transition to a 
ferromagnetic metallic state~\cite{Taniguchi02}, as in the similar compound Sr$_2$RuO$_{4}$ with nearly no octahedral distortion, again indicating that tetragonal splitting is playing a major role in antiferromagnetism. In summary, 
in real materials, as those discussed above, the anisotropy in hoppings and lattice distortions add confusing complexities in the identification of the fundamental properties of the state under scrutiny.

Because at present it is unclear if a condensation of excitons occurs in any transition metal oxide, 
the study of simple idealized models of correlated electrons with SOC that realize this exotic state can help to clarify their basic fundamental properties, removing the distraction created by non-universal lattice distortions.
It is for this reason that here we follow the above proposed route and study the simple degenerate three-orbital Hubbard model and the corresponding large-$U$ spin-orbital model. By performing accurate and unbiased computer simulations of the magnetic excitations, employing state-of-the-art many-body techniques in low dimensions including quantum fluctuations, we managed to establish, and report in this publication, the presence of novel and unique features in the magnetic excitations of the antiferromagnetic excitonic condensate which can be used in both INS and RIXS experiments as {\it fingerprints of excitonic condensates}.


\section{RESULTS}
\noindent {\bf Modelling the excitonic magnet.} 
In this work we will focus on the three-orbital Hubbard model and the effective spin-orbital model 
($SL$ model) with $S=L=1$. The latter can be derived~\cite{Svoboda01} from the $U\gg t,\lambda$ limit 
by perturbation theory.

The three-orbital Hubbard model has three terms: kinetic energy $H_{\textrm{K}}$, on-site Hubbard interactions $H_{\textrm{int}}$, and spin-orbit coupling $H_{\textrm{SOC}}$. The kinetic energy is
\begin{equation}
H_{K} = \sum_{{\langle i ,j \rangle},\sigma,\gamma,\gamma^{\prime}}t_{\gamma\gamma^{\prime}}
(c_{{i}\sigma\gamma}^{\dagger}c^\pdag_{j\sigma\gamma^{\prime}}+\mathrm{H.c.}),
\end{equation}
where $c^{\phantom{\dagger}}_{i\sigma\gamma}(c^{\dagger}_{i\sigma\gamma})$ is the standard fermionic anhilation (creation) operator at site $i$, with spin $\sigma$ and orbital $\gamma$. For simplicity, we consider a hopping unit matrix $t_{\gamma \gamma^{'}}=t\delta_{\gamma\gamma^{'}}$ in orbital space, as in previous studies of excitonic magnetism in the same model~\cite{Kaushal01,Svoboda01,Sato02}. We fix $t=1$ as the unit of energy in the present work.
 
The multiorbital Hubbard interaction consists of four standard components:
\begin{multline}\label{INT_term}
H_{\mathrm{int}} = U\sum_{{i},\gamma} n_{{i}\uparrow\gamma}
n_{{i}\downarrow\gamma} 
+\left(U'-J_{H}/2\right)\sum_{{i},\gamma<\gamma'} n_{{i}\gamma}
n_{{i}\gamma'} 
\\
  -2J_{H}\sum_{{i},\gamma<\gamma'} \mathbf{S}_{{i}\gamma} \cdot 
  \mathbf{S}_{{i}\gamma'} 
+J_{H}\sum_{{i},\gamma<\gamma'} \left( P^{\dagger}_{{i}\gamma} 
P^{\phantom{\dagger}}_{{i}\gamma'} + \mathrm{h.c.} \right),
\end{multline}
where the first and second terms represent the intra-orbital and inter-orbital Coulomb repulsions, respectively. $n_{i\sigma\gamma}$ is the canonical electronic density operator. The third term is the Hund coupling, where $\mathbf{S}_{{i}\gamma}= {{1}\over{2}}\sum_{s,s^{'}} 
c_{{i}s\gamma}^{\dagger} \boldsymbol{\sigma}_{s s^{'}} c^\pdag_{{i}s^{'}\gamma}$ is the spin operator for site $i$ and orbital $\gamma$. In the fourth term $P_{i\gamma}=c_{i\uparrow\gamma}c_{i\downarrow\gamma}$ is the pair anhilation operator, denoting pair-hopping between different orbitals. We fixed $J_{H}=U/4$, as in many recent studies of multiorbital models~\cite{Luo01}, and employ the standard relation $U^{'}=U-2J_{H}$ from rotational invariance.

The on-site SOC term is defined as follows, 
\begin{equation}\label{SO_term}
H_{\mathrm{SOC}}=\lambda\sum_{{i},\gamma,\gamma^{'},\sigma,\sigma^{'}}
{{\langle \gamma|{\bold{L}_{i}}|\gamma^{'}\rangle}\cdot{\langle\sigma|{\bold{S}_{i}}|\sigma^{'}\rangle}}
c_{i\sigma\gamma}^{\dagger}c_{i\sigma^{'}\gamma^{'}} \hspace{0.1cm},
\end{equation} 
where the coupling $\lambda$ is the SOC strength. The electronic density is fixed to 4 electrons per site to represent the $t_{2g}^{4}$ case of our focus.

We also used the $SL$ model, shown below
\begin{multline}\label{SL_model}
H_{SL}=J_{1}^{L}\sum_{<i,j>}\mathbf{L}_{i}\cdot \mathbf{L}_{j} + J_{1}^{S}\sum_{<i,j>}\mathbf{S}_{i}\cdot \mathbf{S}_{j} \\ + J_{2}^{L}\sum_{<i,j>}(\mathbf{L}_{i}\cdot \mathbf{L}_{j})^2 + J_{2}^{SL}\sum_{<i,j>}\mathbf{S}_{i}\cdot \mathbf{S}_{j} \mathbf{L}_{i}\cdot \mathbf{L}_{j} \\
+ J_{3}^{SL}\sum_{<i,j>}\mathbf{S}_{i}\cdot \mathbf{S}_{j} (\mathbf{L}_{i}\cdot \mathbf{L}_{j})^2  + 
\frac{\lambda}{2}\sum_{i}\mathbf{S}_{i}\cdot \mathbf{L}_{i} .
\end{multline}
The details of exchange constants are given in the Methods section. Here we want to point out that: (i) All exchange constants in the $SL$ model are functions of the electron hoping $t$ and interaction strength $U$. (ii) The exchange constants in the above model are positive $J^{\beta}_{\alpha}>0$ ($\beta = L, S, SL$ and $\alpha = 1,2,3$), except $J_{1}^{S}<0$ supporting spin ferromagnetism. (iii) To achieve not only qualitative but also quantitative agreement with the three-orbital Hubbard model, we kept the biquadratic $J_{2}^{L(SL)}$ and triquadratic $J_{3}^{SL}$ terms. (iv) Unless stated otherwise, we will use $U/t=40$ and we will vary the strength of the SOC $\lambda$ parameter in this $SL$ model. As shown below, with such a choice of parameters we find excellent agreement between both models for $U\gg t,\lambda$. It is worth noting that the $SL$ model provides a huge computational benefit for dynamical calculations compared to the three-orbital Hubbard model because of the reduced Hilbert space. The latter also allowed us to obtain results for small two-dimensional clusters using the Lanczos technique.

We solved the above models in one-dimensional geometries using the accurate numerical technique density matrix renormalization group (DMRG)~\cite{White01,White02}. Note that a wide variety of real materials do have dominant 
one-dimensional geometries in their crystal structure, either chains or ladders, 
and recent theoretical investigations have unveiled a wide range of exotic phenomena including block, chiral, superconducting, dimerized, alternating, and Majorana states in one dimensional systems~\cite{1D-block1,1D-block2,1D-block3,1D-block4,1D-ladder1,1D-ladder2,1D-ladder3,1D-ladder4,1D-ladder5,1D-Haldane,1D-Majorana,1D-chiral,1D-dimerized,1D-Gao}. Thus, one dimensionality is not restrictive physically, and allows our study to be
sufficiently accurate to provide reliable many-body information.

\begin{figure}[!b]
\hspace*{-0.52cm}
\vspace*{0cm}
\begin{overpic}[width=1.0\columnwidth , height=1.45\columnwidth]{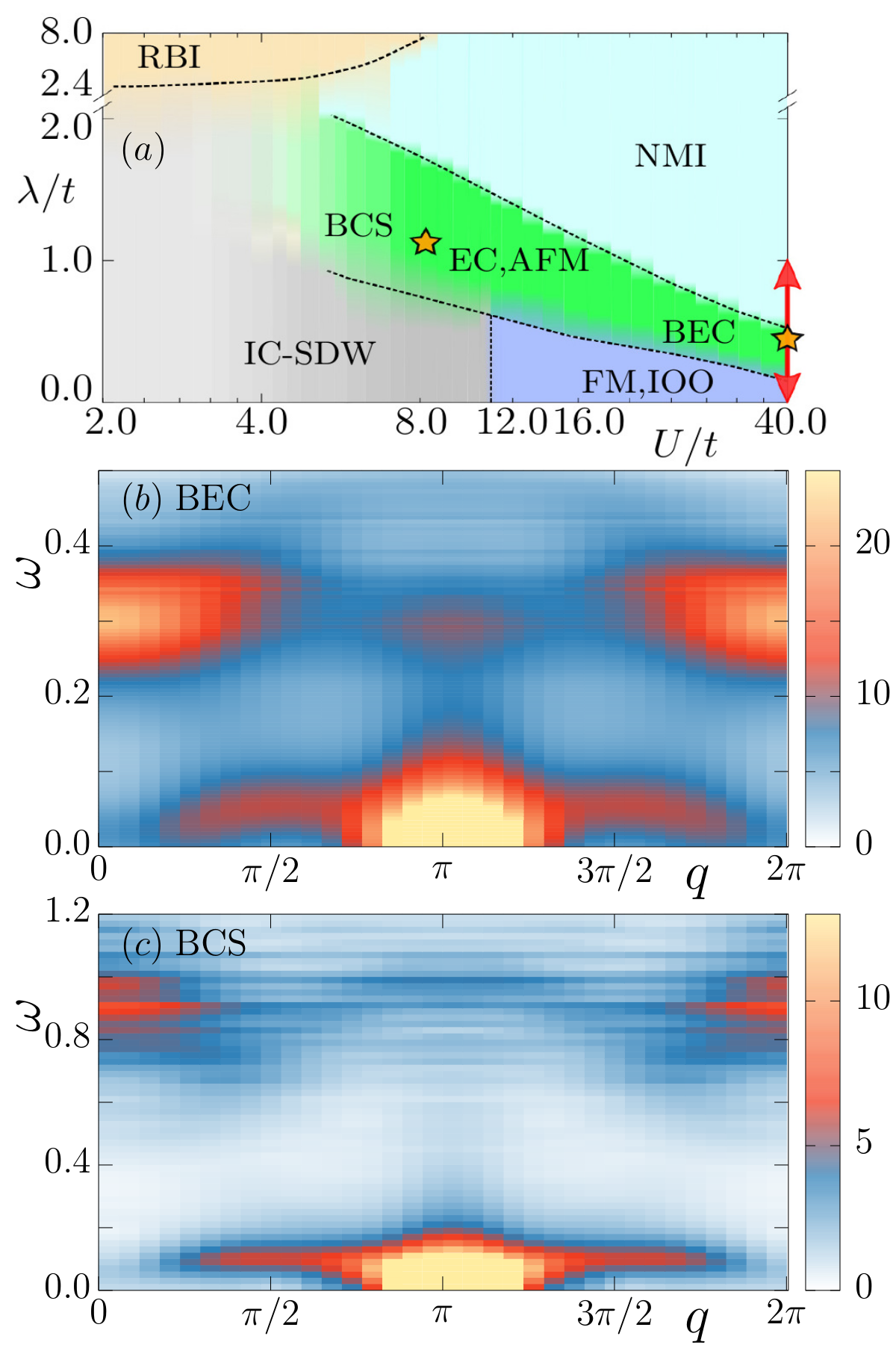}
\end{overpic}
\caption{To guide readers, the $\lambda$ vs $U$ phase diagram is shown in panel (a), reproduced from \cite{Kaushal03}. Panels (b) and (c) show the DMRG dynamical magnetic struture factor $M_{z}(q,\omega)$ for the parameter points $(U/t,\lambda/t)=(40.0,0.45)$ and $(8.0,1.15)$, respectively, depicted as golden stars in (a). Frequency resolution $\Delta\omega/t=0.005$ and $0.02$ is used for (b) and (c), respectively, while fixing the broadening to $\eta/t=0.05$. All above results are calculated using a 16-site three-orbital Hubbard model chain. The red arrow in (a) shows the $\lambda$ range where the large-$U$ effective model was used.}
\label{fig1}
\end{figure}

\vspace{1em}
\noindent {\bf Three-orbital Hubbard model results.} As preliminary, in Fig.~\ref{fig1}(a) we show the $\lambda$ vs $U$ phase diagram obtained using the three-orbital model~\cite{Kaushal03}. A variety of phases are present in Fig.~\ref{fig1}(a): the relativistic band insulator (RBI), incommensurate-spin density wave (IC-SDW), ferromagnetic (FM) phase accompanied with incommensurate orbital ordering (IOO), nonmagnetic insulator (NMI), and the antiferromagnetic (AFM) excitonic condensate (EC) of our focus. We refer the interested readers to Ref.~\onlinecite{Kaushal03} for details of these phases.

The EC region has robust staggered correlations in the triplet channel of the excitons (electron-hole pairs in $j_{\textrm{eff}}=3/2$ and $j_{\textrm{eff}}=1/2$ states), which also leads to antiferromagnetic correlations in both spin and orbital. Furthermore, in recent work investigating the coherence length $\xi$ of exciton pairs, we showed that apart from the short-$\xi$ excitonic Bose-Einstein Condensation (BEC) expected by perturbative calculations at large $U$, the EC also has a Bardeen-Cooper-Schrieffer (BCS) regime at intermediate $U$ with 
intermediate/large $\xi$.

Our goal is to find the answer to the following questions: What kind of magnetic excitations are present in the EC phase of $t_{2g}^4$ systems? Are there unique features (``fingerprints'') of this phase when compared to other possible phases? Clear answers to these questions can be used to identify the $t_{2g}^4$ excitonic condensation when using INS or RIXS experiments. To answer the questions above, first we will discuss the three-orbital Hubbard model dynamical magnetic structure factors in both the BCS and BEC limits of the EC phase, calculated using the formula
\begin{equation}
M_z(q,\omega)=-\frac{1}{\pi}\mathrm{Im}\langle \textrm{gs}|M_{z}(-q) \frac{1}{\omega +i \eta -H +E_{0}} M_{z}(q)|\textrm{gs} \rangle.
\end{equation}
Here $|\textrm{gs}\rangle$ and $E_0$ stand for ground state wave-function and energy, respectively, and $M_{z}(q)$ is the Fourier transform of the local magnetization operator $M_{z}=2S_{z}+L_{z}$ (see also the Methods section). Because the three-orbital does not have any directional anisotropy, here we present only results for the $z$-component of the total moment, but in several cases we verified the transverse components give the same results. Furthermore, because the three-orbital calculations are computationally demanding, further investigation will be done using the large-$U$ $SL$ model. In the latter, we will discuss subtle features of the strong-coupling magnetic excitations and also show the evolution of the spin and orbital excitations tuning $\lambda$ in the range shown in Fig.~\ref{fig1}(a) (red arrow), searching for {\it unique} features in the magnetic excitations to identify the EC phase.

Figure~\ref{fig1}(b) depicts one of the main results of our work, the dynamical magnetic structure factor in the strong-coupling limit, calculated for an $L=16$ site chain and $(U/t,\lambda/t)=(40,0.45)$. The results are dominated by two features: (i) at small frequencies $\omega$, we found a Goldstone-like mode (dispersive spin waves) as expected in an AFM state, with a small bandwidth $\sim 0.06t$, raising from the dominant $q=\pi$ wavevector, a maximum at $q=\pi/2$, and then going down in energy and intensity when approaching $q=0$. (ii) Surprisingly, 
at $\omega/t \sim 0.3$ (several times the spin-wave bandwidth)
we found a gapped optical mode (weakly momentum dependent in energy). The intensity of this mode is largest in the $q=0$ region. Its intensity decreases when increasing $q$, but eventually the intensity increases again near $q=\pi$ leading to a soft patch-like feature at $q=\pi$. The latter mode only appears at small $\xi$ and arises primarily from orbital excitations (as discussed later on).

We also performed similar calculations for $(U/t, \lambda/t)=(8,1.15)$ in the BCS limit, shown in Fig.~\ref{fig1}(c). Similarly as in the BEC regime,
the low-energy spectrum is dominated by a low-energy acoustic branch with bandwidth $\sim 0.1t$. Also as in BEC, we observe an $\omega \sim 0.9t$ optical mode, again with its largest intensity at $q=0$, and this time with no soft patch at $q=\pi$, contrary to BEC. Because of slow convergence issues at large $\omega$, Fig.~\ref{fig1}(c) is more noisy in the gapped mode. Nevertheless, clearly the magnetic excitations in the excitonic phase in both the BCS and BEC limits share common features: a low-energy spin-wave mode, sharpest at $q\sim \pi$, and a high-energy optical mode sharpest at $q \sim 0$.
\begin{figure}[!t]
\hspace*{-0.52cm}
\vspace*{0cm}
\begin{overpic}[width=1.0\columnwidth , height=1.2\columnwidth]{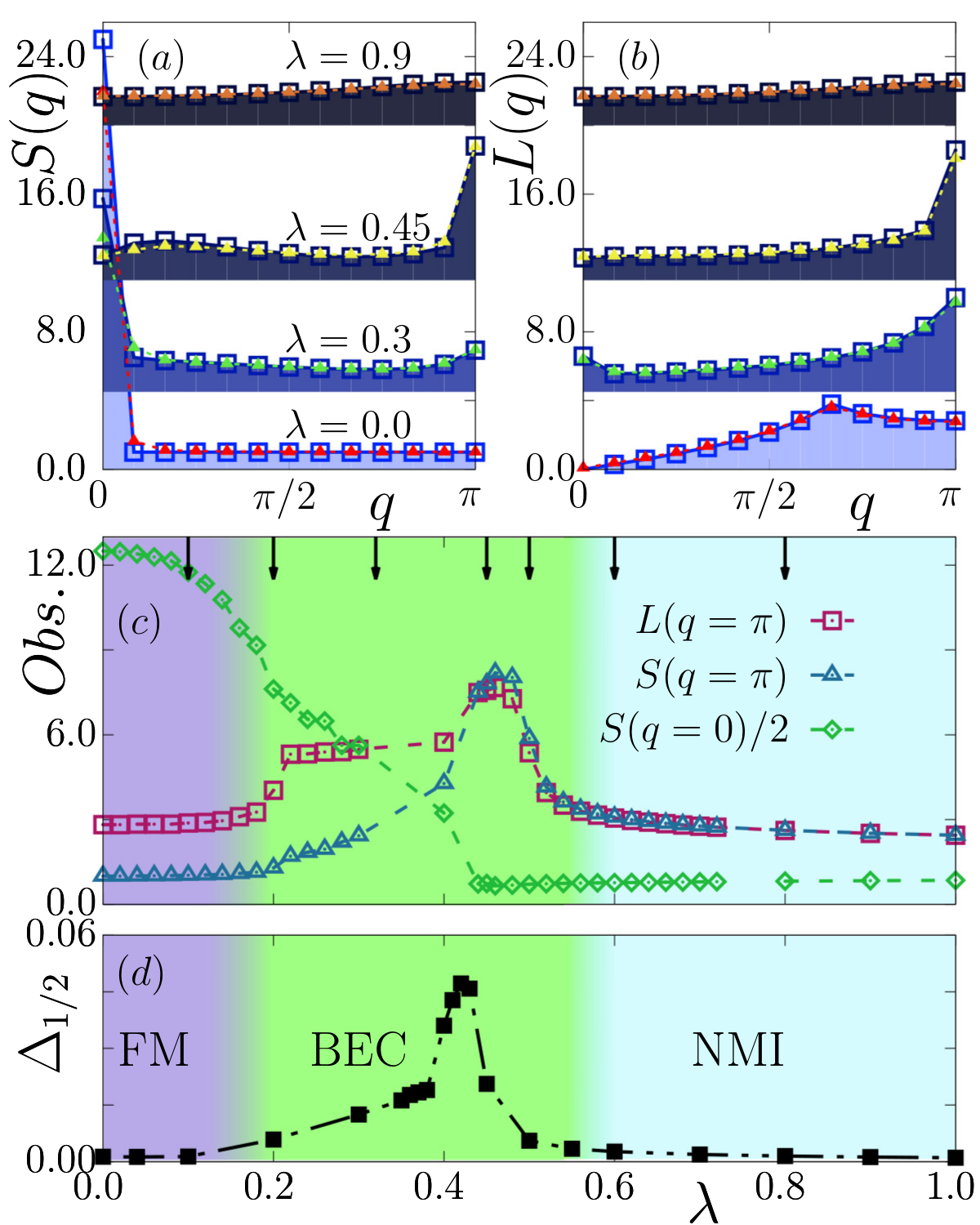}
\end{overpic}
\caption{Spin (a) and orbital (b) static structure factors for the three-orbital Hubbard model (triangles) and the effective spin-orbital $SL$ model (blue filled curves), for various values of $\lambda$. The evolution with $\lambda$ of $S(q=0)$, $S(q=\pi)$, and $L(q=\pi)$ is shown in panel (c), using the $SL$ model. The vertical arrows depict values of $\lambda$ for which dynamical spin and orbital structure factors are presented in Fig.~\ref{fig4}. The excitonic order parameter for the three-orbital Hubbard model is shown in panel (d). For all panels we have used $U/t=40$.}
\label{fig2}
\end{figure}

\vspace{1em}
\noindent {\bf $SL$ model results.} To investigate with better resolution the prominent acoustic and optical modes unveiled in the three-orbital Hubbard model, consider now the strong-coupling $SL$ model.
Figure~\ref{fig2} shows a comparison between the three-orbital Hubbard and $SL$ models at $U/t=40$ and various values of the SOC strength $\lambda$. Panels (a) and (b) depict the static spin $S(q)$ and orbital $L(q)$ structure factors, respectively, as calculated generically using $A(q)=(1/L)\sum_{i,j}e^{\imath q(i-j)}\langle \mathbf{A}_i\cdot \mathbf{A}_j\rangle$, with $\mathbf{A}_i$ either $\mathbf{S}_i$ or $\mathbf{L}_i$. Evidently from the presented results, the $SL$ model captures all the phases extremely  accurately, namely (i) a FM+IOO phase at small $\lambda$, with incipient AFM tendecies (strong peak at $q=0$ and weak peak at $q=\pi$) at $\lambda/t=0.3$, (ii) the excitonic condensate AFM phase at $\lambda/t=0.45$, and (iii) the non-magnetic insulating phase at $\lambda/t=0.9$, the latter showing no order as reflected by flat static structure factors.
 
Figure~\ref{fig2}(c) displays the evolution of $S(0)$, $S(\pi)$, and $L(\pi)$ with $\lambda/t$. At $\lambda<0.19t$ the system is FM, while AFM tendencies start above $\lambda/t=0.19$. The latter is accompanied by staggering in excitons as measured by $\Delta_{m}=\frac{1}{L^{2}} \sum_{|i-i'|>0}(-1)^{|i-i'|}\langle \Delta_{1/2,m}^{\dagger 3/2,m}(i) \Delta_{1/2,m}^{3/2,m}(i') \rangle$ in the three-orbital model, 
as shown in Fig.~\ref{fig2}(d), where $\Delta_{1/2,m}^{\dagger 3/2,m}(i)$ is an exciton creation operator at site $i$ (for details see Ref.~\cite{Kaushal03}). Both the AFM and excitonic order are maximized near $\lambda \sim 0.45t$. For $\lambda/t>0.55$ the system becomes nonmagnetic.
\begin{figure*}[!t]
\hspace*{-0.cm}
\vspace*{0cm}
\begin{overpic}[width=2.0\columnwidth , height=0.9\columnwidth]{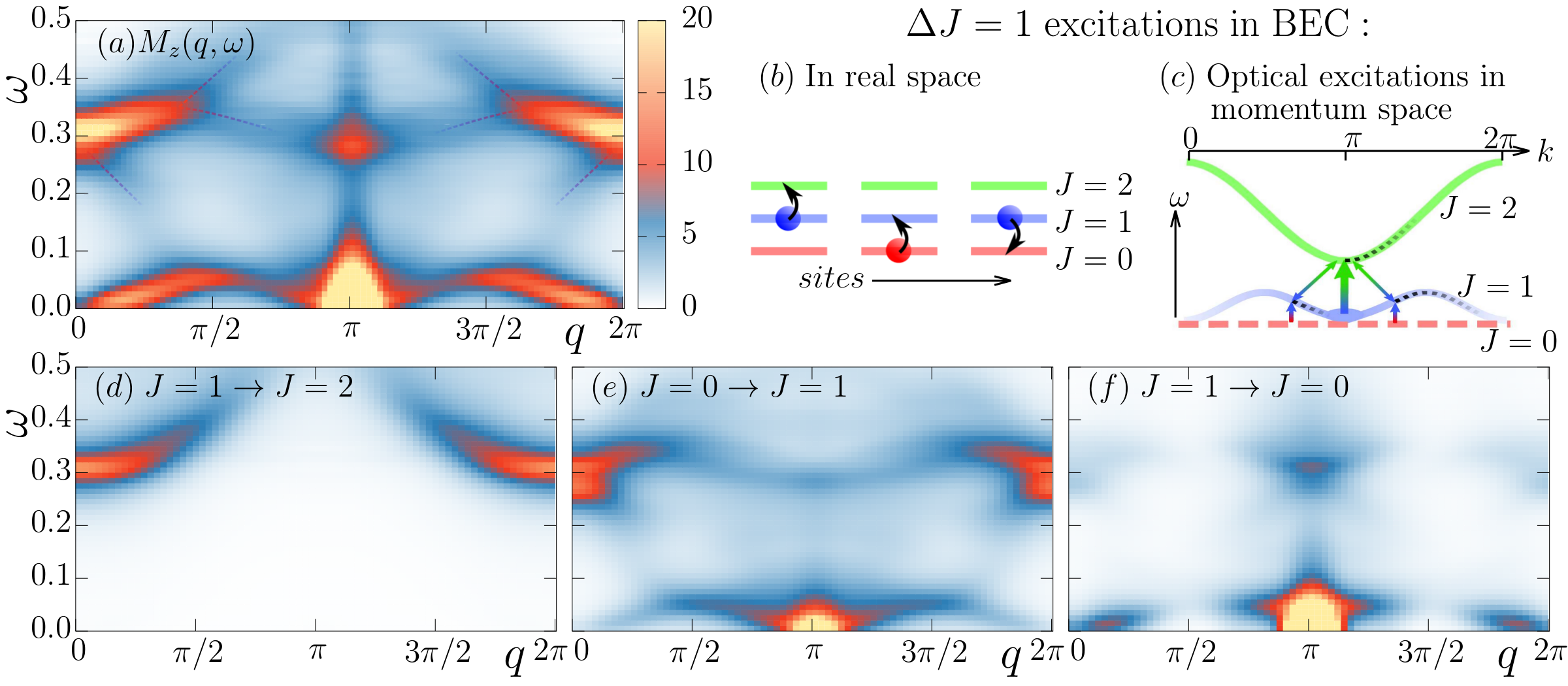}
\end{overpic}
\caption{(a) Dynamical magnetic structure factor $M_{z}(q,\omega)$, obtained with DMRG, $L=48$ sites, and the $SL$ model. (b) shows a cartoon illustrating the allowed $\Delta J=1$ excitations on the ground state of the BEC phase. (c) describes visually the optical magnetic excitation, using $J=1$ and $J=2$ spin-orbit excitons in momentum space. (d,e,f) Dynamical structure factors calculated for the specific excitation channels shown in panel (b), using $L=36$ sites chain. For all the results above $\lambda/t=0.45$, $U/t=40$, $\Delta\omega/t=0.01$, and $\eta/t=0.02$ were used.}
\label{fig3}
\end{figure*}

Consider now the magnetic excitations of the $SL$ model at $(U/t,\lambda/t)=(40,0.45)$ shown in Fig.~\ref{fig3}(a), where the excitonic magnetic phase is the strongest. Because of the better momentum resolution of the $SL$ model compared to the three-orbital model, now more subtle features of the optical mode near $\omega/t \sim 0.3$ become visible. As discussed earlier (using the three-orbital model), the optical excitation is well defined at $q \sim 0$. The advantage of the $SL$ model is that now
three decaying branches become prominently revealed
moving away from $q=0$ (dashed lines are guides to the eyes). A high intensity patch at $q \sim \pi$ near $\omega/t \sim 0.3$ is also present, as in the three-orbital results. 

\begin{figure}
\hspace*{-0.52cm}
\vspace*{0cm}
\begin{overpic}[width=1.0\columnwidth , height=2.0\columnwidth]{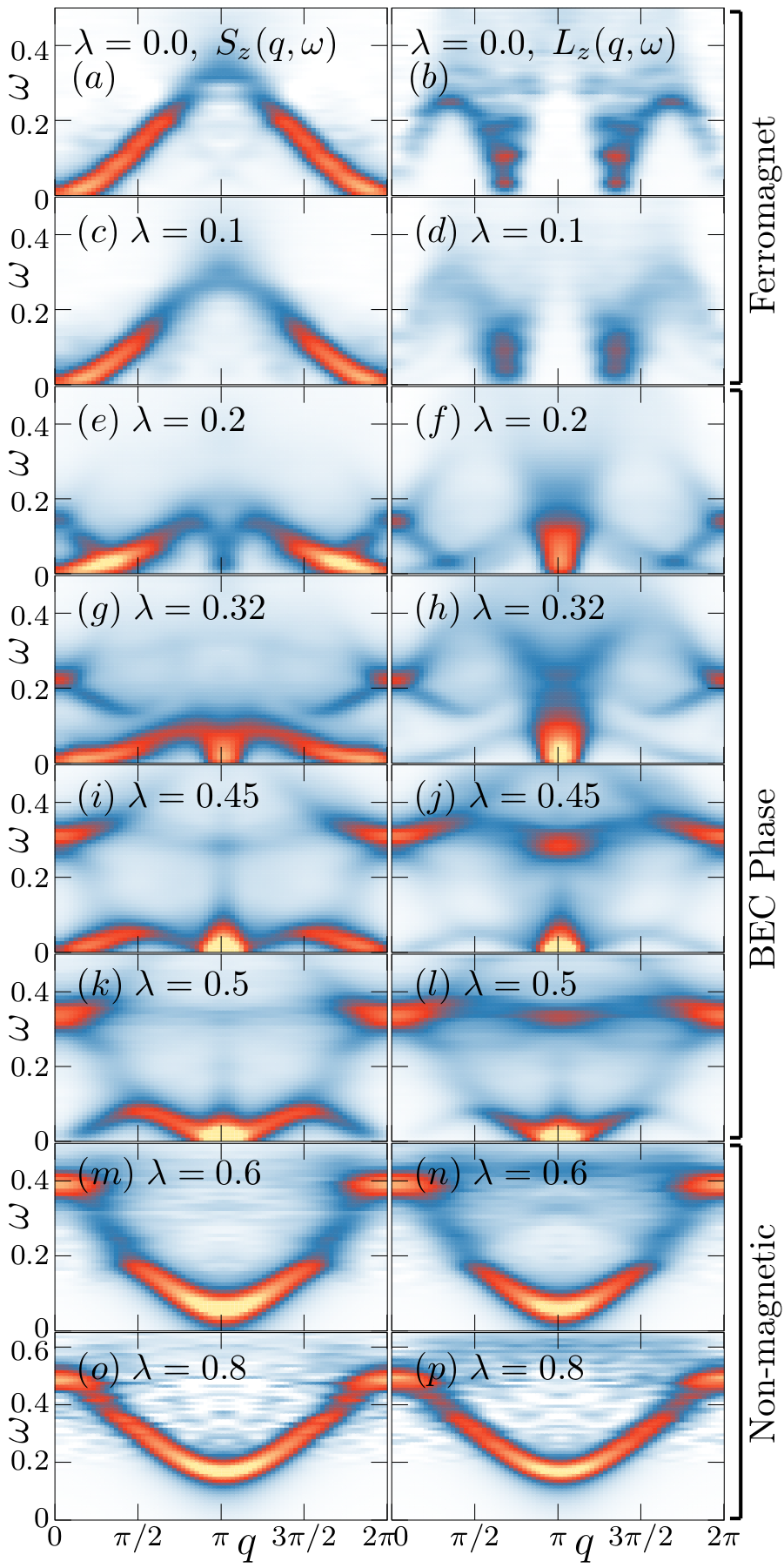}
\end{overpic}
\caption{Panels on the left and right hand side show the evolution of the dynamical spin structure factor $S_{z}(q,\omega)$ and dynamical orbital structure factor $L_{z}(q,\omega)$ of the $SL$ model, respectively, with increasing $\lambda/t$. An $L=36$ chain was used for all results, at fixed $U/t=40.0$, $\Delta\omega/t=0.01$, and $\eta/t=0.02$.}
\label{fig4}
\end{figure}

To better understand the optical excitations, we performed Lanczos studies on small chain clusters of $L=4$ and 6 sites. This allows for an intuitive analysis of the eigenstates contributing to the optical mode. Remarkably, even in these small Lanczos systems similar optical modes as in DMRG were clearly found. From the exact ground state of those clusters, we observed that primarily two excited states contribute to the optical mode, denoted by $|O_{1}^{q=0}\rangle$ and $|O_{2}^{q=0}\rangle$. 
Expanding $|\textrm{gs}\rangle$ and those two excited states $|O_{1}^{q=0}\rangle$ and $|O_{2}^{q=0}\rangle$ in the local $|J,J_{z}\rangle$ basis (the natural basis at strong $\lambda/t$) yields  (at $\lambda/t=0.45$ and $L=6$) 
\begin{equation}
|\textrm{gs}\rangle = \sum_{j}c_{gs}e^{i\pi j} ... |0,0\rangle_{j-1} |1,0 \rangle_{j}|0,0\rangle_{j+1}... + ....
\end{equation}
with dominant $L|c_{gs}|^{2}\approx 0.62$ over other weights (the triplon is in only one site $j$). 
The above wave-function indicates that 
the ground state represents one triplon with momentum $\pi$. Similarly, we found the excited states are dominated by two terms in the basis used, namely 
\begin{multline}
|O_{l}^{q=0}\rangle=\sum_{j}c_{l}^{1}e^{i\pi j} ...|0,0\rangle_{j-1}|2,0\rangle_{j}|0,0\rangle_{j+1}... +
\\
  \sum_{j}c_{l}^{2}e^{i\pi j} ...|1,0\rangle_{j}|0,0\rangle_{j+1}|1,0\rangle_{j+2}... + ....
\end{multline}
where the first term has a quintuplon only at site $j$, while the second term has two triplons at $j$ and $j+2$ sites.
Regarding their weights
$|c_{1}^{1}|\approx 0.303$, $|c_{1}^{2}|\approx 0.105$, $|c_{2}^{1}|\approx 0.127$, and $|c_{2}^{2}|\approx 0.208$ so that $L(|c_{l}^{1}|^2 + |c_{l}^{2}|^2)$ are $\approx 0.62$ and $0.35$ for $l=1$ and $2$, respectively. The comparable values of coeffcients $|c_{l}^{j}|$, for given $l$, shows that both the $J=2$ and $J=1$ states are involved in the optical modes in the EC. The above analysis also suggests that the $q=0$ optical excitation predominantly arises from the (i) $J=1$ to $J=2$ and (ii) $J=0$ to $J=1$ transitions. Furthermore, our Lanczos results indicate that the simple singlet-triplet models (ignoring $J=2$ states), used to analyze the EC spin-wave spectrum~\cite{Akbari01} cannot properly describe the optical excitations discussed here (though they can still correctly predict the ground state). Investigations at smaller $\lambda/t$'s in the BEC phase, where the ground state has multiple triplons, yields similar conclusions, further supporting the origin of the optical excitations. 

To confirm the above scenario using larger lattices (available to the DMRG method), we have evaluated the susceptibility of the $\Delta J=1$ transitions, defined as
\begin{equation}
P^{JJ^{'}}(q,\omega)=-\frac{1}{\pi}\mathrm{Im}\langle \mathrm{gs}|(P^{JJ^{'}}_{q})^{\dagger} \frac{1}{\omega +i \eta -H +E_{0}} P^{JJ^{'}}_{q}|\mathrm{gs}\rangle\,,
\end{equation}
where $P^{JJ^{'}}_{q}$ is the Fourier transform of operators which excite the site $i$ from the state $|J^{'},0\rangle_{i}$ to the state $|J,0\rangle_{i}$, i.e., $P^{JJ^{'}}_{i}=|J,0\rangle_{i}\langle J^{'},0|_{i}$. In our analysis we focused on the transition suggested by Lanczos results (\mbox{$J=1\to 2$} and \mbox{$J=0\to 1$}), and also on \mbox{$J=1\to 0$} since the symmetric operator $M_{z}$ in the $|J,J_{z}\rangle$ basis connects such states as well. Although the \mbox{$J=2\to 1$} excitations are also allowed by the selection rules, we ignore them, since $J=2$ states have a vanishing contribution to the ground state in the excitonic condensate BEC regime. 

The dynamical susceptibilities are shown in Figs.~\ref{fig3}(d,e,f). Our results indicate, in agreement with conclusions reached from Lanczos data, that the \mbox{$J=1\to 2$} excitation clearly relates to the top-most branch of the optical mode, while the other two branches of the optical mode arise from the \mbox{$J=0\to 1$} excitations. Furthermore, we found that the slightly less intense patch at $q\sim\pi$ near $\lambda\simeq0.3t$ arises from the $J=1\to 0$ excitation. Finally, only the excitations between $J=0$ and $J=1$ contribute toward the acoustic mode. 

We can understand the momentum dependence of the optical excitations by analyzing the spin-orbit excitonic excitations (\mbox{$J=0\to 1$} and \mbox{$J=0\to 2$}) in momentum space. In the NMI state, it is known that both \mbox{$J=0\to 1$} and \mbox{$J=0\to 2$} spin-orbit excitons have cosine-like bands with minima at momentum $k=\pi$ and a gap $\lambda$ between $J=1$ and $J=2$ states~\cite{Svoboda01}.
In the BEC phase, we found that the \mbox{$J=0\to 1$} excitation is heavily renormalized due to triplon condensation, i.e. the acoustic mode in Fig.~\ref{fig3}(e), and it is safe to assume that \mbox{$J=0\to 2$} excitations has still a cosine-like band because high-energy quintuplons are not important in the condensation of triplons. These heauristic arguments lead to the sketch presented in Fig.~\ref{fig3}(c). Now it is clear that excitations of triplon with momentum $k=\pi$ to quintuplon with momentum $k=\pi$ should have $\Delta k=q=0$ and cosine dependence on $q$, explaining the optical excitation in Fig.~\ref{fig3}(d). This excited $J=2$ state can decompose into two triplons, as shown in Fig.~\ref{fig3}(c), which leads to two more optical branches following the dispersion of the acoustic branch but at higher energy because the $J=2$ state is involved. This explains the presence of both optical branches of $J=0$ to $J=1$ excitations in Fig.~\ref{fig3}(e), i.e., (i) the optical branch with non-monotonic dispersion, extending up to $q=\pi$, and (ii) the other optical branch with energy decreasing as $q$ increases.

\vspace{1em}
\noindent {\bf SOC dependence.} Figure~\ref{fig4} shows the evolution with $\lambda/t$ of individual contributions to the magnetic excitations, namely the  dynamical spin and orbital structure factor, $S_{z}(q,\omega)$ and $L_{z}(q,\omega)$, respectively. In the FM phase [$\lambda/t=\{0.0,0.1\}$, panels (a-d)], outside the excitonic condensate, there are spin waves in $S_z(q,\omega)$ emerging from $q=0$, as expected for a canonical FM state. $L_{z}(q,\omega)$ is also gapless, and has a maximum intensity close to $q=2\pi/3$. The presence of a broad continuum in $L_{z}(q,\omega)$ at elevated $\omega$ indicates that the system is an orbital spin liquid state. In the $\lambda\to0$ limit, the $SL$ model can be written as a bilinear-biquadratic Hamiltonian, for which fractionalization with ordering vector $q=2\pi/3$ was recently observed~\cite{Feng01, Feng02}. Interestingly, the FM spin waves lose intensity in $(\omega,q)$-space where a continuum of orbital excitations exist.

For $\lambda/t\gtrsim0.2$, in the region where excitons acquire staggering order [panels (e,f)], we observe the development of the $q\to0$ optical mode, in both the spin and orbital excitations.
Furthermore, the orbital excitations $L_{z}(q,\omega)$ are heavily renormalized, forming gapless excitations at $q=\pi$ with large intensity. 
Further increasing $\lambda/t$ inside the BEC region - panels (e-n) for $\lambda/t=\{0.20,0.32,0.45,0.50\}$ - increases the energy of the $q=0$ optical mode. Simultaneously,  the spin acoustic mode, $S_{z}(q\to\pi,\omega\to0)$, increases in intensity, while the FM contribution $S_{z}(q\to0,\omega\to0)$ decreases. The complexity of the acoustic mode can be attributed to the competition between FM spin exchange, in the absence of SOC, and the AFM tendencies due to the staggered excitons in the EC region. It is interesting to note that the $q=\pi$ high-energy mode is mainly captured by orbital excitations, see panels (j,l).

Finally, we also present results in the nonmagnetic phase at large SOC [$\lambda/t={0.6, 0.8}$, panels (m-p)]. Here, both spin and orbital excitations are gapped and now the optical mode's lower branch is connected with the acoustic mode leading to the formation of a single band corresponding to the \mbox{$J=0\to 1$} excitation. 
This is consistent with the nonmagnetic phase ground state being predominantly a direct product of local singlets, $|GS\rangle_{NMI} \approx |0,0\rangle_{1}|0,0\rangle_{2}|....|0,0\rangle_{L}$, and hence $J=0$ to $J=1$ transitions primarily contribute to the magnetic excitations. 

\begin{figure}[!t]
\hspace*{-0.52cm}
\vspace*{0cm}
\begin{overpic}[width=1.0\columnwidth , height=1.5\columnwidth]{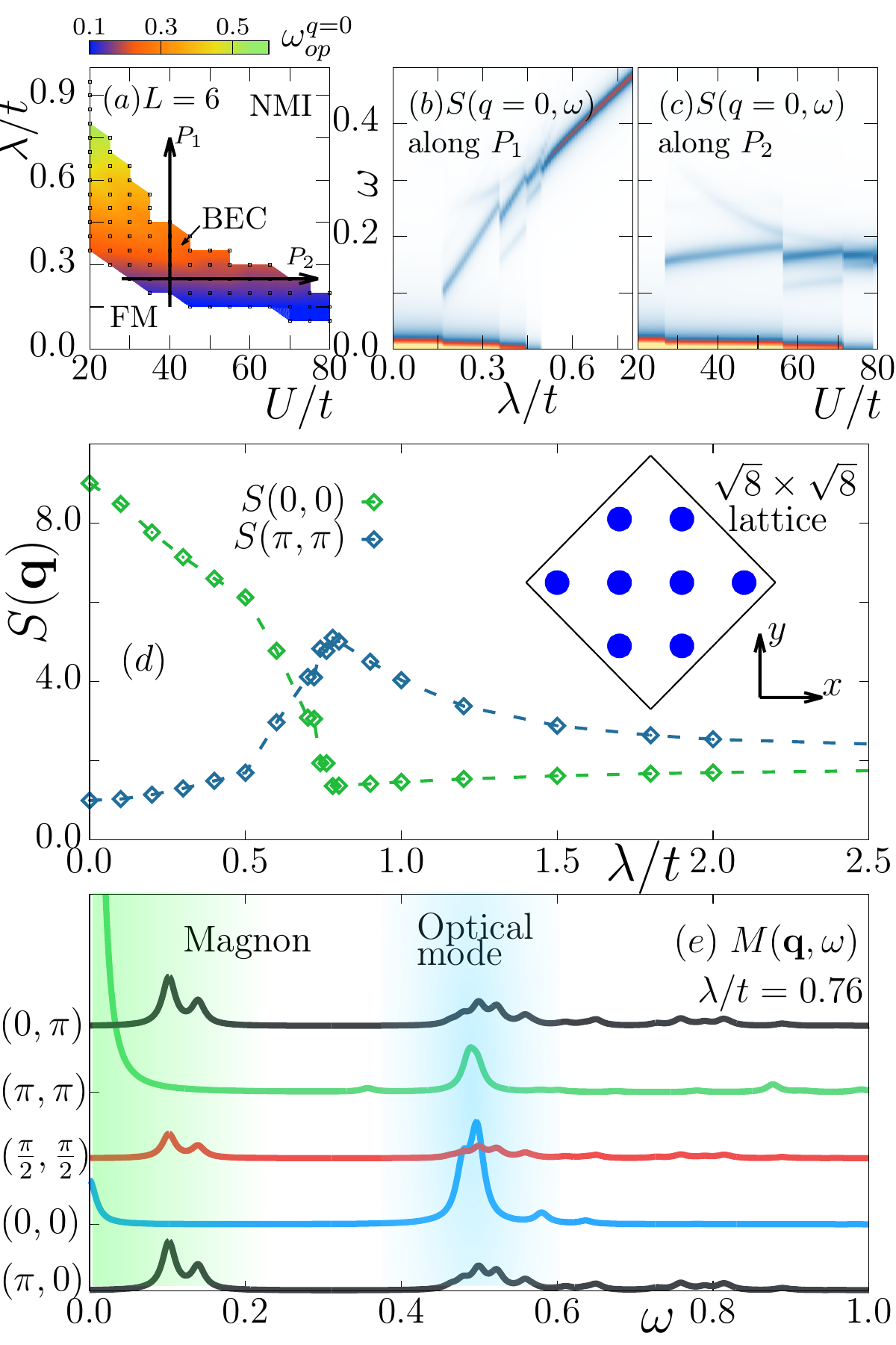}
\end{overpic}
\caption{(a) shows the $\lambda/t$-$U/t$ phase diagram calculated via the Lanczos method on an $L=6$ chain and using the $SL$ model. Colors inside the BEC phase are choosen following the optical excitation energy $\omega_{op}^{q=0}$. Panels (b,c) show the evolution of the long-wavelength dynamical spin structure factor, $S(q=0,\omega)$, for the $P_{1(2)}$ paths shown in panel (a). The small discontinuities visible in the plot are caused by changes in the total $J$ of the ground state by one unit.  
 Panels (d,e) depict Lanczos results for an $\sqrt{8}\times \sqrt{8}$ cluster at fixed $U/t=40$ and $J_{H}=U/4$. In (d), $S(\mathbf{q})$ for various $\lambda/t$'s, at the dominant momenta $\mathbf{q}=(0,0)$ and $(\pi,\pi)$, are shown. The cluster used in Lanczos analysis appears as inset. (e) Dynamical magnetic structure factor $M(\mathbf{q},\omega)$ at $\lambda/t=0.76$, with magnon and optical modes similar to the one dimensional case.}
\label{fig5}
\end{figure}

To unveil what energy scale dominates the optical mode position, we calculated $S(q,\omega)$ with Lanczos using a $L=6$ chain at various $\lambda/t$'s and $U/t$'s. To help readers, in Fig.~\ref{fig5}(a)  we show the BEC region in the $\lambda/t$-$U/t$ phase diagram of the $SL$ model. Colors are chosen based on the energy $\omega_{op}^{q=0}$ of the $q=0$ optical excitation. We use two paths: $P_{1}$ at constant $U/t=40$ (Figs.~\ref{fig5}(b)) and $P_{2}$
at constant $\lambda/t=0.25$ (Figs.~\ref{fig5}(c)).
Evidently, the optical mode energy increases approximately linearly with $\lambda/t$ (roughly $\approx 0.67\lambda/t$), whereas it is rather independent of $U/t$. These findings are consistent with our previously discussed qualitative description of the optical mode as an intra-site excitation. In strong coupling, $U/t$ only enters via the exchange terms. Note that the $J=1 \to 2$ excitation energy is only mildly renormalized from $\lambda$ (in the pure atomic limit) to nearly $0.67\lambda$.

\vspace{1em}
\noindent {\bf Two-dimensions.} We complete our study discussing the dependence of our results on the dimensionality of the lattice. Using the Lanczos method, we studied the $8$-site tilted square cluster [see inset of Fig.~\ref{fig5}(d)], widely employed in other contexts because it preserves the $D_{4h}$ symmetry of the square lattice. In Fig.~\ref{fig5}(d), fixing $U/t=40$, $J_{H}=U/4$, and tuning $\lambda/t$, we show the evolution of the spin static structure factors $S(\mathbf{q})$ at momenta $\mathbf{q}=(0,0)$ and $(\pi,\pi)$. The transition from FM to AFM order, and eventually to a non-magnetic state with increasing $\lambda/t$, are similar as observed in one-dimensional chains [compare with Fig.~\ref{fig2}(c)]. Figure~\ref{fig5}(e) shows the dynamical $M(\mathbf{q},\omega)$ at $\lambda/t=0.76$, where $S(\pi,\pi)$ is maximized and hence provides a good representative point for the BEC excitonic magnet phase. Interestingly, similarly as in chains, we again found a dominant low-energy AFM acoustic (magnon) mode, with highest intensity at $\mathbf{q}=(\pi,\pi)$, a small-intensity $\mathbf{q}=(0,0)$ peak caused by the FM-phase vicinity, and, more importantly, a high-energy optical mode at $\omega/t \approx 0.5$ with largest weight at $\mathbf{q}\to(0,0)$. The above results show that the long-wavelength optical mode in the BEC phase is not limited to chains but extends to planes, and can be used as a fingerprint of the excitonic condensation in quasi two-dimensional $t_{2g}^4$ materials as well.

\section{Discussion}

We have presented a comprehensive study of the dynamical magnetic structure factors of the one-dimensional three-orbital Hubbard and effective $SL$ models, as well as results on small two-dimensional clusters, in the regime where the SOC creates an excitonic magnet. We found unique features in the magnetic excitations of the $t_{2g}^{4}$ excitonic condensate: (1) An acoustic magnetic wave emerging from $q=\pi$, as expected from an AFM magnetic system, coexisting with small but non-zero weight close to $q=0$, the latter remnant of the $\lambda\to0$ case. (2) A novel optical mode with sharp intensity in the $q\to 0$ limit, notorious in our results. Using the effective $SL$ model we showed that the optical mode originates from the $J=1\to2$ and $J=0\to1$ transitions. Our Lanczos study on small systems and DMRG investigation on larger systems, has shown that the coexistence of both features -- distorted spin wave and optical mode -- are uniquely associated to the excitonic magnet by comparing to other phases in the phase diagram. As a consequence, such features can be used as a fingerprints of the excitonic magnet in the analysis of INS or RIXS experimental data for $t_{2g}^4$ materials.

A gapped high-energy mode was also observed in the magnetic excitations of Sr$_2$IrO$_4$ ($t_{2g}^5$) in RIXS experiments~\cite{JKim01}. This excitonic band of Sr$_2$IrO$_4$ is the excitation of an electron from a low-energy $j_{\textrm{eff}}=3/2$ band (fully occupied by electrons) to a high-energy $j_{\textrm{eff}}=1/2$ band (half-filled occupied). However, in this case the ground state is a $j_{\textrm{eff}}=1/2$ Mott insulator, not an excitonic condensate.  
This RIXS spin-orbit exciton can be described as a doped single hole (or electron) in a half-filled one-orbital Hubbard model. The reason is that the excited electron in the $j_{\textrm{eff}}=1/2$ band and the hole in the $j_{\textrm{eff}}=3/2$ band form a local bound state~\cite{JKim02}. 

The optical band unveiled in our $t_{2g}^{4}$ study cannot be described by a similar rationale as in the $t_{2g}^5$ case. In an excitonic magnet, the spin-orbit excitons (triplons) are {\it already preformed} in the ground state, and the excitations correspond to the annihilation of a $J=1$ triplon (i.e. a spin-orbit exciton) followed by the creation of a $J=2$ quintuplon (another spin-orbit exciton). The quintuplon excitation can split into two triplons, by the action of the Hamiltonian, but they are still triplons interacting with one another. Hence, the state with interacting triplons can be captured by singlon-to-triplon excitations as well. 
In simpler terms, while in the $t_{2g}^5$ case the optical mode is the creation of an individual exciton, starting with no excitons in the ground state, the $t_{2g}^4$ optical mode reported here
represents a $J=1$ exciton to $J=2$ exciton transition. This can only happen when $J=1$ excitons are already present in the ground state, as in the EC ground state of our focus.

Recent RIXS experiments on Ca$_2$RuO$_4$~\cite{Gretarsson01} reported a high-energy optical mode at energy $\approx 0.32eV$ and attributed it to $J=0 \to 2$ excitations using single-atom states. However, due to strong tetragonal splitting in Ca$_2$RuO$_4$, the gap between $J=1 \to 2$ is of similar size~\cite{Gretarsson01}. Assuming Ca$_2$RuO$_4$ has triplon condensation in the ground state, as suggested in various works~\cite{Jain01,Akbari01,Feld01,Lotze01,Strobel}, we may expect a $\Delta J=1$ optical mode near the same energy, measurable by INS experiments. Finding this optical mode by increasing the experimental energy range may help to clarify the microscopic understanding of the ground state of Ca$_2$RuO$_4$.

Furthermore, our findings may be of direct relevance for one-dimensional $t_{2g}^{4}$ materials, such as  $\textrm{OsCl}_{4}$~\cite{Cotton01} and $\textrm{Sr}_{3}\textrm{NaIrO}_{6}$~\cite{Ming01}. Especially the later is of interest, because a recent experimental study~\cite{Bandyopadhyay01} showed the presence of short-range antiferromagnetic correlations and gapless spin excitations resembling the antiferromagnetic excitonic condensate phase. Interestingly, one-dimensional stripes can also be engineered starting from higher dimensional compounds, as recently shown for $\textrm{Sr}_{2}\textrm{IrO}_{4}$ ($t_{2g}^{5}$)~\cite{Gruenewald01}. Our study should further encourage experimentalists to synthesize other low-dimensional $t_{2g}^{4}$ materials and investigate their magnetic excitations with INS or RIXS searching for the above reported fingerprints of an exciton magnet.

\section{Methods}

\noindent {\bf DMRG method.} For the three-orbital Hubbard model we used the corrected single-site DMRG algorithm with correction $a$=0.001~\cite{White03}, keeping up to 1500 states and performing 35 finite sweeps for proper convergence. We applied the two-site DMRG for the $SL$ model, keeping up to 1500 states and maintaining the truncation error below $10^{-6}$. For the main results of this work, i.e. the dynamical magnetic structure factors, we used the correction vector method dynamical DMRG~\cite{Kuhner01}, assisted by the Krylov-space approach~\cite{Nocera01}. 

\noindent {\bf Exchange constants in the $SL$ model.} As described in the main text, the $SL$ model can be derived from the three-orbital model in the large-$U$ limit~\cite{Svoboda01}. In our study we fixed $J_{H}=U/4$ and $U^{'}=U-2J_{H}$, so all the exchange constants can be written in term of $t^{2}/U$ as:  $J_{1}^{L}=5t^{2}/U$, $J_{1}^{S}=-4t^{2}/U$, $J_{2}^{L}=31t^{2}/6U$, $J_{2}^{SL}=3t^{2}/U$, and $J_{3}^{L}=17t^{2}/6U$.
Because in the $SL$ model the effective $L$ operator used is the negative of the $L$ operator in the three orbital model, the operator ${\bf{M}}=2{\bf S}-{\bf L}$ is used for the $SL$ model calculations.

\noindent{\bf Details of magnetic excitations calculations.}
We used the Krylov-space based DMRG-correction vector target method to calculate the dynamical structure factors $A(q,\omega)$ defined as 
\begin{equation}
A(q,\omega)=\frac{1}{L}\sum_{i,j} e^{\iota q (j-i)} A_{ij}(\omega).
\label{EqAqw}
\end{equation}

In the equation above, $A_{ij}(\omega)$ is defined as
\begin{equation}
A_{ij}(\omega)=-\frac{1}{\pi}\mathrm{Im}\langle gs|A^{\dagger}_{j} \frac{1}{\omega +i \eta -H +E_{0}} A_{i}|gs \rangle,
\end{equation}
where the operator $A$ can be $S_{z}, \; L_{z}, \; M_{z}, {\rm or} \; P^{JJ^{'}}$. 
To reduce computational cost, in Eq.~(\ref{EqAqw}) we fixed site $i=L/2$ at the center of the chain and instead used
\begin{equation}
A(q,\omega)=\frac{1}{\sqrt{L}}\sum_{j} e^{\iota q (j-L/2)} A_{iL/2}(\omega).
\end{equation}
Furthermore, we also used the parzen filter to reduce the finite-size effects~\cite{Kuhner01}.
To reproduce the data shown in this publication, the free and open source DMRG++ program~\cite{alvarezDMRGpp} and input files are available at {\it https://g1257.github.io/dmrgPlusPlus/}.

\section{Acknowledgements}
N.~K. and E.~D. were supported by the US Department of Energy (DOE), Office of Science, Basic Energy Sciences (BES), Materials Science and Engineering Division. J.~H. acknowledges grant support by the Polish National Agency for Academic Exchange (NAWA) under Contract No.~PPN/PPO/2018/1/00035 and by the National Science Centre (NCN), Poland, via Project No.~2019/35/B/ST3/01207. G.~A. was partially supported by the Scientific Discovery through Advanced Computing (SciDAC) program funded by U.S. DOE, Office of Science, Advanced Scientific Computing Research and BES, Division of Materials Sciences and Engineering. The work of G.~A. was conducted at the Center for Nanophase Materials Science, sponsored by the Scientific User Facilities Division, BES, DOE, under contract with UT-Battelle.

\section{Author contributions}
N.K., J.H., and E.D. planned the project. N.K. performed the calculations. 
G.A. developed the \textsc{DMRG++} computer program. N.K, J.H., G.A. and E.D. 
wrote the manuscript. All co-authors provided comments on the paper.

\section{Additional information}
\noindent {\bf Competing Interests} The authors declare no competing interests.

\FloatBarrier

\end{document}